\documentclass{IEEEtran}
\usepackage{array} 
\usepackage{graphicx} 
\usepackage{subcaption} 
\usepackage{amsmath} 
\usepackage{amsfonts} 
\usepackage{cite}
\usepackage{caption}
\usepackage{balance}
\usepackage[ruled,vlined,linesnumbered,ruled]{algorithm2e}
\usepackage{lettrine} 
\usepackage{hyperref} 
\hypersetup{hypertex=true,
	colorlinks = true,
	linkcolor = blue,
	anchorcolor = blue,
	citecolor = blue}
\begin{document}
	
\title{Deep Reinforcement Learning for Energy Efficiency Maximization in RSMA-IRS-Assisted ISAC System}

\author{Zhangfeng Ma, \emph{IEEE Member}, Ruichen Zhang, \emph{IEEE Member}, Bo Ai, \emph{IEEE Fellow}, Zhuxian Lian, \\
             Linzhou Zeng, \emph{IEEE Member}, and Dusit Niyato, \emph{IEEE Fellow}

\thanks{

Z. Ma and B. Ai are with the  School of Electronic and Information Engineering, Beijing Jiaotong University, Beijing 100044, China (e-mails: zhangfeng.ma@vip.126.com, boai@bjtu.edu.cn). R. Zhang and D. Niyato are with the College of Computing and Data Science, Nanyang Technological University, Singapore 639798 (e-mail: ruichen.zhang@ntu.edu.sg, dniyato@ntu.edu.sg). Z. Lian and L. Zeng are with the School of Information and Communications Engineering, Faculty of Electronic and Information Engineering, Xi'an Jiaotong University, Xi'an 710049, China (e-mails: zhuxianlian@just.edu.cn, linzhou.zeng@hnist.edu.cn).

}

}
\maketitle
\thispagestyle{empty}

\begin{abstract}
	
This paper proposes a three-dimensional (3D) geometry-based channel model to accurately represent intelligent reflecting surfaces (IRS)-enhanced integrated sensing and communication (ISAC) networks using rate-splitting multiple access (RSMA) in practical urban environments. Based on this model, we formulate an energy efficiency (EE) maximization problem that incorporates transceiver beamforming constraints, IRS phase adjustments, and quality-of-service (QoS) requirements to optimize communication and sensing functions. To solve this problem, we use the proximal policy optimization (PPO) algorithm within a deep reinforcement learning (DRL) framework. Our numerical results confirm the effectiveness of the proposed method in improving EE and satisfying QoS requirements. Additionally, we observe that system EE drops at higher frequencies, especially under double-Rayleigh fading.

\end{abstract}


\section{Introduction}

\lettrine[lines=2]{S}{ince} the integrated sensing and communication (ISAC) combines communication and sensing capabilities, it enables concurrent data transmission and environmental monitoring, which significantly improves resource utilization and reduces operational complexity \cite{Ref1}. However, high path loss and blockage probability in line-of-sight (LoS) scenarios limit its practical implementation. To mitigate these challenges, intelligent reflecting surfaces (IRS) employ reflective elements to redirect electromagnetic waves towards the desired direction, enhancing sensing performance \cite{Ref2}, \cite{Ref3}. Meanwhile, as the complexity of resource sharing and signal processing increases in ISAC systems, traditional multiple access schemes may struggle to meet system demands, leading to higher energy consumption \cite{Ref4}. Fortunately, rate-splitting multiple access (RSMA) can be introduced to optimize user data splitting and interference management, thereby improving energy utilization \cite{Ref5}. Therefore, the integration of IRS and RSMA offers strong support for the evolution of ISAC systems.

So far, only a limited number of researchers have focused on optimizing the performance of IRS-assisted ISAC systems with RSMA \cite{Ref6, Ref7, Ref8}. In \cite{Ref6}, a novel alternating optimization method was applied to maximize the achievable unicast rate, demonstrating the superiority of this resource allocation strategy under the perfect successive interference cancellation (SIC) scenario. In \cite{Ref7}, an algorithm was introduced to optimize EE while considering constraints on user service quality and the sensing signal-to-noise ratio (SNR), showing significant improvements in EE. In \cite{Ref8}, an iterative algorithm was devised to maximize the SNR of the target detection. However, these studies primarily relied on the ideal channel model, which assumes time-invariant conditions and omits Doppler effects between transceivers. Thus, the insights derived may lack both accuracy and rigor.

Motivated by the above, we study the transmission design for RSMA-IRS-assisted ISAC system under practical channel fading conditions. The major contributions of this paper are as follows. \emph{{\textbf{Firstly}}}, we establish a geometry-based channel model, which consists of a dual-functional base station (BS), multiple communication users, and a moving target (i.e. unmanned aerial vehicle). Moreover, the proposed model has the ability to reflect the different fading effects by adjusting model parameters such as distance, velocity and angle information. \emph{\textbf{Secondly}}, we formulate a transmission design strategy for EE maximization is formulated, considering various quality of service (QoS) requirements, transceiver beamforming and practical phase shifts. Then, we adopt a proximal policy optimization (PPO) algorithm based on deep reinforcement learning (DRL) theory is adopted to tackle the formulated optimization problem. \emph{\textbf{Thirdly}}, simulation results show that the proposed algorithm significantly outperforms those based on traditional space-division multiple access (SDMA). Furthermore, the impacts of carrier frequency and the radar cross section (RCS) area on the EE are both discussed.

\section{System Model and Problem Formulation}

\subsection{System Model}

\begin{figure}[!t]
\centering{\includegraphics[width=0.48\textwidth]{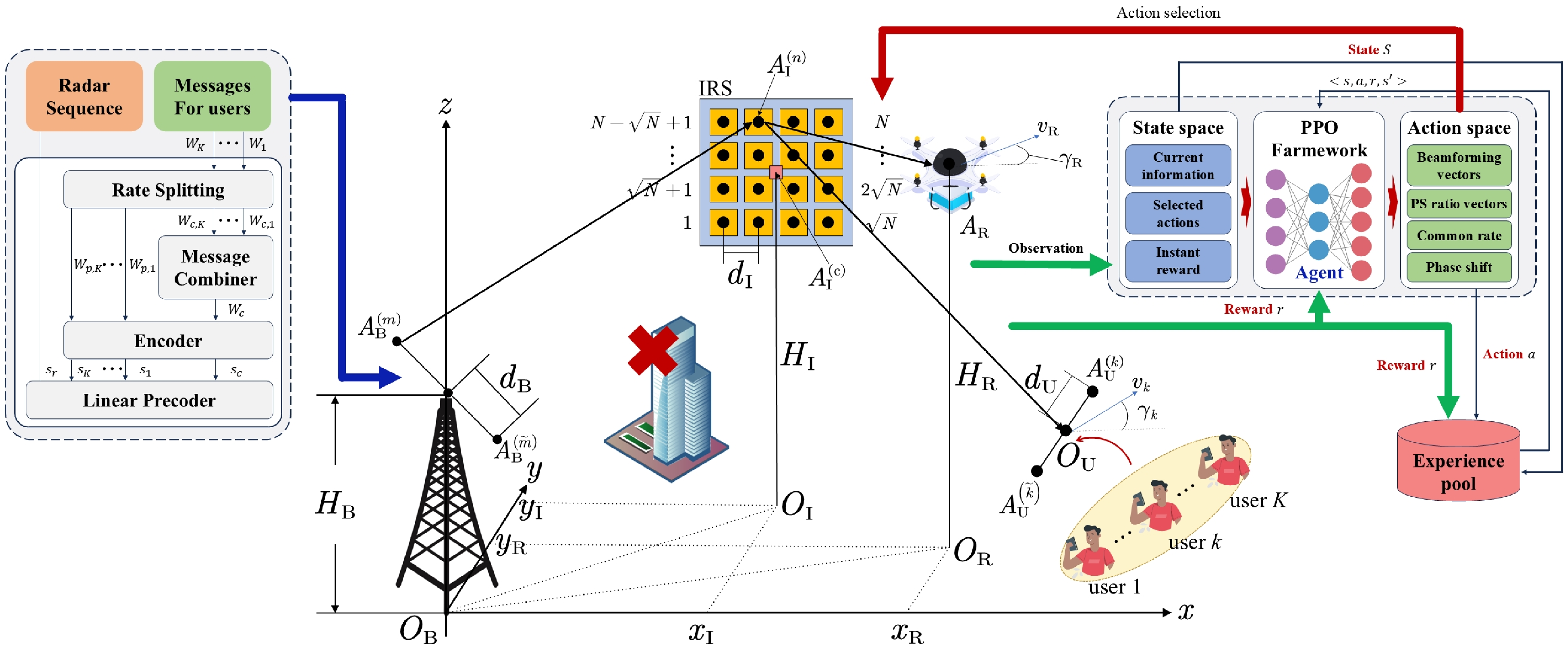}}
\caption{System model of an IRS-enhanced ISAC network with RSMA. In this setup, user messages are split into common and private parts for efficient interference management. The PPO-based DRL framework is used to enhance the system EE and meet QoS requirements.}
\label{Fig1}
\end{figure}

In this paper, we consider an IRS-assisted ISAC system as illustrated in Fig. \ref{Fig1}, where a dual-function BS simultaneously transmits information to $K$ single-antenna communication users, and detects a target. The $K$ users collectively form a uniform linear array (ULA), in which each element represents a single-antenna user. Furthermore, we consider that the BS is equipped with ULA consisting of $M$ elements. Given that the direct links from the BS to the users/target are obstructed by scatterers (e.g., buildings, humans and trees), the IRS is deployed on a high-rise building to improve the quality of ISAC services. Fig. \ref{Fig1} shows that the IRS is equipped with a square planar array consisting of $\sqrt N \times \sqrt N$ elements, arranged in $\sqrt N$ rows along the $x$-axis and $\sqrt N $ columns along the $z$-axis, with equal inter-element spacing $d_I$. Furthermore, the sets of BS antenna elements, users, and IRS elements are denoted by $m \in \mathcal{M} = \left\{ {1, \cdots ,M} \right\}$, $k \in \mathcal{K} = \left\{ {1, \cdots ,K} \right\}$, and $n \in \mathcal{N} = \left\{ {1, \cdots ,N} \right\}$, respectively. The centers of the BS, IRS, and MS antennas are located at ${O_{\rm{I}}}$, ${O_{\rm{T}}}$, and ${O_{\rm{U}}}$, respectively, with corresponding coordinates $\left( {{x_{\rm{I}}},{y_{\rm{I}}}} \right)$, $\left( {{x_{\rm{T}}},{y_{\rm{T}}}} \right)$, and $\left( {{x_{\rm{U}}},{y_{\rm{U}}}} \right)$. The inter-element spacing for the BS and user antenna arrays is denoted by ${d_{\rm{B}}}$ and ${d_{\rm{U}}}$, respectively. The $k$th user moves with velocity ${\upsilon _k}$ at an angle ${\gamma _k}$, while the radar target moves with velocity ${\upsilon _{\rm{R}}}$  at an angle ${\gamma _{\rm{R}}}$. In coordinate plane, the 3D components of the $m$th BS antenna element $A_{\rm{B}}^{\left( m \right)}$, the $k$th user $A_{\rm{U}}^{\left( k \right)}$ and the radar target ${A_{\rm{R}}}$ can be written as $A_{\rm{B}}^{\left( m \right)} \buildrel \Delta \over = \left( {0,{{\left( {M - 2m + 1} \right){\delta _{\rm{B}}}} \mathord{\left/{\vphantom {{\left( {M - 2m + 1} \right){\delta _{\rm{B}}}} 2}} \right. \kern-\nulldelimiterspace} 2},{H_{\rm{B}}}} \right)^{\rm{T}}, \forall m \in \mathcal{M}$, $A_{\rm{U}}^{\left( k \right)}\buildrel \Delta \over = {\left( {{x_{\rm{U}}},{{\left( {K - 2k + 1} \right){\delta _{\rm{U}}}} \mathord{\left/{\vphantom {{\left( {K - 2k + 1} \right){\delta _{\rm{U}}}} 2}} \right.\kern-\nulldelimiterspace} 2},0} \right)^{\rm{T}}}, \forall k \in \mathcal{K}$, ${A_{\rm{R}}} \buildrel \Delta \over = \left( {{x_{\rm{R}}},{y_{\rm{R}}},{H_{\rm{R}}}} \right)^{\rm{T}}$, where ${H_{\rm{B}}}$ and ${H_{\rm{R}}}$ are the heights of the BS and target, respectively. Moreover, the 3D components of the $n$th IRS element $A_{{\rm{I}}}^{\left( n \right)}$  and the center of the IRS ${A_{\rm{I}}^{\left( {\rm{c}} \right)}}$ can be written as $A_{\rm{I}}^{\left( n \right)}{\rm{ }} \buildrel \Delta \over = {\left( {{x_{\rm{I}}} + {\Delta _n},{y_{\rm{I}}},{H_{\rm{I}}} + {\Delta _n}} \right)^{\rm{T}}}, \forall n \in {\cal N}$, ${A_{\rm{I}}^{\left( {\rm{c}} \right)}} \buildrel  \Delta \over = {\left( {{x_{\rm{I}}} ,{y_{\rm{I}}},{H_{\rm{I}}} } \right)^{\rm{T}}}$, where ${\Delta _n} = \left( {n - \left\lfloor {{{\left( {n - 1} \right)} \mathord{\left/{\vphantom {{\left( {n - 1} \right)} {\sqrt N }}} \right.\kern-\nulldelimiterspace} {\sqrt N }}} \right\rfloor \cdot\sqrt N  - \mathcal{G} } \right){\delta _{\rm{I}}}$, $\mathcal{G}  = \left\lfloor {{{\left( {\sqrt N  + 1} \right)} \mathord{\left/{\vphantom {{\left( {\sqrt N  + 1} \right)} 2}} \right.\kern-\nulldelimiterspace} 2}} \right\rfloor  + 0.5 \cdot \bmod \,\left( {\sqrt N  + 1,2} \right)$, and ${H_{\rm{I}}}$ is the height of the center of the IRS.

It can be observed from Fig. \ref{Fig1} that there are mainly two types of the links in the system, i.e., the BS-IRS-user channel and the BS-IRS-target channel. Specifically, the channel impulse response (CIR) from the $m$th transmit antenna to the $n$th IRS element is represented by ${\bf{G}}= {\left[ {{g_{mn}}} \right]^{\rm{T}} \in \mathbb{C} ^{N \times M}}$, while CIRs from the $n$th IRS element to the $k$th user and the radar target are respectively represented by ${{\bf{h}}_k}\left( t \right) = {\left[ {{h_{1k}}\left( t \right), \cdots ,{h_{nk}}\left( t \right), \cdots ,{h_{Nk}}\left( t \right)} \right]}^{\rm{T}} \in {\mathbb{C} ^{N \times 1}}$ and ${{\bf{h}}_{r}}\left( t \right) = {\left[ {{h_{1{r}}}\left( t \right), \cdots ,{h_{n{r}}}\left( t \right), \cdots ,{h_{N{r}}}\left( t \right)} \right]}^{\rm{T}} \in {{\mathbb{C} }^{N \times 1}}$, where ${g_{mn}}$, ${h_{nk}}\left( t \right)$, ${h_{n{\rm{R}}}}\left( t \right)$ are respectively given by
\begin{equation}\label{eq1}
{g_{mn}} = \frac{\lambda }{{4\pi {\varepsilon _{A_{\rm{B}}^{\left( {m} \right)}A_{\rm{I}}^{\left( {n} \right)}}}}}\left( {\bar g_{mn}^{{\rm{BI}}} + \tilde g_{mn}^{{\rm{BI}}}} \right),
\end{equation}
\begin{equation}\label{eq2}
{h_{nk}}\left( t \right) = \frac{\lambda }{{4\pi {\varepsilon _{A_{\rm{I}}^{\left( {n} \right)}A_{\rm{U}}^{\left( k \right)}}}}}\left( {\bar h_{nk}^{{\rm{IU}}}\left( t \right) + \tilde h_{nk}^{{\rm{IU}}}\left( t \right)} \right),
\end{equation}
\begin{equation}\label{eq3}
{h_{n{r}}}\left( t \right) = \sqrt {\frac{{{\lambda ^2}\sigma }}{{{{\left( {4\pi } \right)}^3}{{\left( {{\varepsilon _{A_{\rm{I}}^{\left( {n} \right)}{A^{\left( {r} \right)}}}}} \right)}^4}}}} \left( {\bar h_{n{r}}^{{\rm{IR}}}\left( t \right) + \tilde h_{n{r}}^{{\rm{IR}}}\left( t \right)} \right),
\end{equation}
where ${\lambda} = {{c_0} \mathord{\left/{\vphantom {{c_0} f_c }} \right.\kern-\nulldelimiterspace} f_c}$ is wavelength; $f_c$ is the carrier frequency; $c_0$ designates the speed of light; $\sigma$ represents the RCS of the target. Correspondingly, the LoS components in (\ref{eq1})--(\ref{eq3}) are respectively written as
\begin{equation}\label{eq4}
\bar g_{mn}^{{\rm{BI}}} = \sqrt {\frac{{{K_{{\rm{BI}}}}}}{{{K_{{\rm{BI}}}} + 1}}} \exp \left( {{{ - j2\pi \sqrt {{\varepsilon _{A_{\rm{B}}^{\left( m \right)}A_{\rm{I}}^{\left( m \right)}}}} } \mathord{\left/
 {\vphantom {{ - j2\pi \sqrt {{\varepsilon _{A_{\rm{B}}^{\left( m \right)}A_{\rm{I}}^{\left( m \right)}}}} } \lambda }} \right.
 \kern-\nulldelimiterspace} \lambda }} \right),
\end{equation}
\begin{eqnarray}\label{eq5}
\begin{split}
\bar h_{nk}^{{\rm{IU}}}\left( t \right) &= \sqrt {\frac{{{K_{{\rm{IU}}}}}}{{{K_{{\rm{IU}}}} + 1}}} \exp \left( {{{ - j2\pi \sqrt {{\varepsilon _{A_{\rm{I}}^{\left( n \right)}A_{\rm{U}}^{\left( k \right)}}}} } \mathord{\left/
{\vphantom {{ - j2\pi \sqrt {{\varepsilon _{A_{\rm{I}}^{\left( n \right)}A_{\rm{U}}^{\left( k \right)}}}} } \lambda }} \right.\kern-\nulldelimiterspace} \lambda }} \right) \\& \times  \exp \left( {j2\pi t{f_{{\rm{IU}},{\rm{LoS}}}}} \right),
\end{split}
\end{eqnarray}
\begin{eqnarray}\label{eq6}
\begin{split}
\bar h_{nr}^{{\rm{IR}}}\left( t \right) &= \sqrt {\frac{{{K_{{\rm{IR}}}}}}{{{K_{{\rm{IR}}}} + 1}}} \exp \left( {{{ - j2\pi \sqrt {{\varepsilon _{A_{\rm{I}}^{\left( n \right)}{A^{\left( r \right)}}}}} } \mathord{\left/
{\vphantom {{ - j2\pi \sqrt {{\varepsilon _{A_{\rm{I}}^{\left( n \right)}{A^{\left( r \right)}}}}} } \lambda }} \right.\kern-\nulldelimiterspace} \lambda }} \right) \\& \times \exp \left( {j2\pi t{f_{{\rm{IR}},{\rm{LoS}}}}} \right),
\end{split}
\end{eqnarray}
where $K_{{\rm{BI}}}$, $K_{{\rm{IU}}}$ and $K_{{\rm{IR}}}$ denote the Rician factors for the BS-IRS channel, IRS-user and IRS-target channel, respectively. The propagation distance terms  in (\ref{eq4})--(\ref{eq6})  can be expressed as ${\varepsilon _{A_{\rm{B}}^{\left( m \right)}A_{\rm{I}}^{\left( n \right)}}} = {{{\left( {{x_{mn}}} \right)}^2} + {{\left( {{y_{mn}}} \right)}^2} + {{\left( {{z_{mn}}} \right)}^2}}$, ${\varepsilon _{A_{\rm{I}}^{\left( n \right)}A_{\rm{U}}^{\left( k \right)}}} = {{{\left( {{x_{nk}}} \right)}^2} + {{\left( {{y_{nk}}} \right)}^2} + {{\left( {{z_{nk}}} \right)}^2}}$, ${\varepsilon _{A_{\rm{I}}^{\left( n \right)}{A^{\left( {{r}} \right)}}}} =  {{{\left( {{x_{n{{r}}}}} \right)}^2} + {{\left( {{y_{n{{r}}}}} \right)}^2} + {{\left( {{z_{n{{r}}}}} \right)}^2}}$, where ${x_{mn}} = {x_{\rm{I}}} + {\Delta _n}$, ${y_{mn}} = {y_{\rm{I}}} - {{\left( {M - 2m + 1} \right){d_{\rm{B}}}} \mathord{\left/{\vphantom {{\left( {M - 2m + 1} \right){d_{\rm{B}}}} 2}} \right.\kern-\nulldelimiterspace} 2}$, ${z_{mn}} = {H_{\rm{I}}} + {\Delta _n} - {H_{\rm{B}}}$, ${x_{nk}} = {x_{\rm{I}}} + {\Delta _n} - {x_{\rm{U}}}$, ${y_{nk}} = {y_{\rm{I}}} - {{\left( {K - 2k + 1} \right){\delta _{\rm{U}}}} \mathord{\left/{\vphantom {{\left( {K - 2k + 1} \right){\delta _{\rm{U}}}} 2}} \right.\kern-\nulldelimiterspace} 2}$, ${z_{nk}} = {H_{\rm{I}}} + {\Delta _n}$, ${x_{n{r}}} = {x_{\rm{I}}} + {\Delta _n} - {x_{r}}$, ${y_{n{r}}} = {y_{\rm{I}}} - {y_{r}}$, ${z_{n{\rm{r}}}} = {H_{\rm{I}}} + {\Delta _n} - {H_{r}}$. The Doppler terms in (\ref{eq5}) and (\ref{eq6}) can be expressed as
\begin{equation}\label{eq7}
{f_{{\rm{IU}},{\rm{LoS}}}} = \frac{{{\upsilon _k}}}{\lambda }\frac{{\left( {{x_{\rm{I}}} - {x_{\rm{U}}}} \right)\cos {\gamma _k} + {y_{\rm{I}}}\sin {\gamma _k}}}{{\sqrt {{{\left( {{x_{\rm{I}}} - {x_{\rm{U}}}} \right)}^2} + {{\left( {{y_{\rm{I}}}} \right)}^2}} }},
\end{equation}
\begin{equation}\label{eq8}
{f_{{\rm{IR}},{\rm{LoS}}}} = \frac{{{\upsilon _r}}}{\lambda }\frac{{\left( {{x_{\rm{I}}} - {x_r}} \right)\cos {\gamma _r} + \left( {{y_r} - {y_{\rm{I}}}} \right)\sin {\gamma _r}}}{{\sqrt {{{\left( {{x_{\rm{I}}} - {x_r}} \right)}^2} + {{\left( {{y_{\rm{I}}} - {y_r}} \right)}^2}} }}.
\end{equation}
Finally, the corresponding non-LoS (NLoS) components of the channel in (\ref{eq1})--(\ref{eq3}) can be respectively modeled by $\tilde h_{nm}^{{\rm{BI}}} \sim{\cal CN}\left( 0,1 \right)$, $\tilde h_{nm}^{{\rm{IU}}} \sim{\cal CN}\left( 0,1 \right)$ and $\tilde h_{nm}^{{\rm{IR}}} \sim{\cal CN}\left( 0,1 \right)$.

Unlike existing works on ISAC, the 1-layer RSMA scheme is employed at the BS to serve multiple communication users. Specifically, the message associated with the $k$-th user ${W_k}{\left( t \right)}$ at time $t$ is split into two parts, i.e., the common part $W_k^{\left( {\rm{c}} \right)}{\left( t \right)}$ and the private part $W_k^{\left( {\rm{p}} \right)}{\left( t \right)}$. Then the common parts of all users' messages $W_k^{\left( {\rm{c}} \right)}{\left( t \right)}$ are combined into a public message $W^{\left( {\rm{c}} \right)}{\left( t \right)}$, which is encoded into the common stream $s_{\rm{c}}{\left( t \right)}$ using a codebook shared by all users. Each private part $W_k^{\left( {\rm{p}} \right)}{\left( t \right)}$, containing the remaining parts of the messages $W_k{\left( t \right)}$, is independently encoded into the private stream $s_k{\left( t \right)}$ for $k$th user. Accordingly, the transmit signal vector is given by
\begin{eqnarray}\label{eq9}
{\bf{x}}\left( t \right) = \underbrace {\overbrace {{{\bf{v}}_c}\left( t \right){{{s}}_c}\left( t \right)}^{{\rm{Common}}\;{\rm{stream}}} + \overbrace {\sum\limits_{k \in \mathcal{K}} {{{\bf{v}}_k}\left( t \right){{{s}}_k}\left( t \right)} }^{{\rm{Private}}\;{\rm{streams}}}}_{{\rm{Communication}}\;{\rm{streams}}} + \underbrace {{{\bf{v}}_r}\left( t \right){{{s}}_r}\left( t \right)}_{{\rm{Radar}}\;{\rm{stream}}},
\end{eqnarray}
where ${\bf{v}}{\left( t \right)} = \left[ {{{\bf{v}}_{c}}{\left( t \right)},{{\bf{v}}_1\left( t \right)},\cdots,{{\bf{v}}_K}{\left( t \right)},{{\bf{v}}_r}{\left( t \right)}} \right] \in {\mathbb{C}^{M \times (K + 2)}}$ is the transmit beamforming matrix. Specifically, ${{\bf{v}}_{c}}{\left( t \right)}\in {\mathbb{C}^{M}}$, ${{{{\bf{v}}_k}\left( t \right)}}\in {\mathbb{C}^{M}}$, and ${{\bf{v}}_{r}}{\left( t \right)}\in \mathbb{C}^{M}$ are respectively the beamformer of the common stream, private stream, radar stream. ${\bf{s}}{\left( t \right)} = \left[ {{{{s}}_{c}}{\left( t \right)},{{{s}}_1\left( t \right)},\cdots,{{{s}}_K}{\left( t \right)},{{{s}}_r}{\left( t \right)}} \right] ^{\rm{T}}\in {\mathbb{C}^{(K + 2) \times 1}}$ is the transmit stream, where $\mathbb{E}\left\{ {{{\bf{s}}}{\bf{s}}^H} \right\} = {{\bf{I}}_{K + 2}}$. ${{s}_{c}}{\left( t \right)}$ is the common stream, ${{{{{s}}_k}\left( t \right)}}$ is the private stream, and ${{{s}}_{r}}{\left( t \right)}$ is the radar sequence. At the $k$th user, the received signal is given by
\begin{equation}\label{eq10}
{y_k}\left( t \right) = {{\bf{F}}_k}\left( t \right){\bf{x}}\left( t \right) + {n_k}\left( t \right),
\end{equation}
where ${{\bf{F}}_k}\left( t \right) = {\bf{h}}_k^H\left( t \right){\bf{\Phi }}\left( t \right){\bf{G}}$ denotes the cascaded channel for user $k$. ${\bf{\Phi }}\left( t \right) = {\rm{diag}}\left( {\exp \left( {j{\phi _1{\left( t \right)}}} \right), \cdots ,\exp \left( {j{\phi _N{\left( t \right)}}} \right)} \right)\in \mathbb{C} ^{N\times N}$ denotes the phase shift matrix, ${\phi _n} \in \left[ {0,2\pi } \right)$ is the phase shift of the $n$th unit of the IRS. $n_{k} {\left( t \right)}\sim \mathcal{C} \mathcal{N}\left(0, \delta _{k}^{2}\right)$ represents the additive white Gaussian noise (AWGN) at user $k$. At the BS, the radar echo, ${{{\bf{y}}_r}}\in \mathbb{C} ^{M\times 1}$, for the sensing functionality is given by
\begin{equation}\label{eq11}
{{\bf{y}}_r}\left( t \right) = {\bf{{\hat F}}}\left( t \right){{\bf{v}}_r}\left( t \right){s_r}\left( t \right) + {{\bf{n}}_r}\left( t \right),
\end{equation}
where ${\bf{{\hat F}}}\left( t \right) = {{\bf{G}}^H}{\bf{\Phi }}\left( t \right){{\bf{h}}_r}\left( t \right){\left( {{{\bf{h}}_r}\left( t \right)} \right)^H}{\left( {{\bf{\Phi }}\left( t \right)} \right)^H}{\bf{G}}$ denotes the equivalent channel of the detection signal. ${{\bf{n}}_r}\left( t \right)\sim \mathcal{C} \mathcal{N}\left(0, \delta _{r}^{2}{{\bf{I}}_M}\right)$ denotes the AWGN received at the BS. In the following discussion, the variable $\left( t \right)$ is omitted for brevity. The EE of the proposed system can be expressed as
\begin{equation}\label{eq12}
\eta  = {R \mathord{\left/{\vphantom {R P}} \right.\kern-\nulldelimiterspace} P},
\end{equation}
where $P = \mu \left( {{{\left\| {{{\bf{v}}_{\rm{c}}}} \right\|}^2} + \sum\limits_{k \in {\cal K}} {{{\left\| {{{\bf{v}}_k}} \right\|}^2}}  + \chi  \cdot {{\left\| {{{\bf{v}}_{\rm{r}}}} \right\|}^2}} \right) + {P_{{\rm{ST}}}}$ denotes the transmit power; the binary variable $\chi$ is determined by the user's ability to SIC of the radar sequence. If the SIC of $s_r$ is possible, $\chi = 0$; otherwise, $\chi = 1$; $\mu  \in \left[ {1, + \infty } \right)$ denotes the power amplifier efficiency factor; ${{P_{{\rm{ST}}}}}$ denotes the static hardware power. $R$ denotes the sum achievable rate, i.e., $R = \sum\limits_{k \in \mathcal{K}} {\left( {{C_k} + R_k^{\left( {\rm{p}} \right)}} \right)}$, where $C_k$ is the allocated common rate to $k$-th user. It satisfies that $\mathop {\min }\limits_k \left\{ {\left. {R_k^{\left( {\rm{c}} \right)}} \right|k \in {\cal K}} \right\} \ge \sum\limits_{k \in {\cal K} } {{C_k}}$, where $R_k^{\left( {\rm{c}} \right)} = {\log _2}\left( {1 + {\rm{SINR}}_k^{\left( {\rm{c}} \right)}} \right)$ denotes the achievable rate for the common stream at $k$-th user, ${{\rm{SINR}}_k^{\left( {\rm{c}} \right)}}$ represents the corresponding signal-to-interference-plus-noise (SINR), i.e.,
\begin{equation}\label{eq13}
{{\rm{SINR}}_k^{\left( {\rm{c}} \right)} = \frac{{{{\left| {{{\bf{F}}_k}{{\bf{v}}_c}} \right|}^2}}}{{\sum\limits_{i \in {\cal K}} {{{\left| {{{\bf{F}}_k}{{\bf{v}}_i}} \right|}^2}}  + \chi {{\left| {{{\bf{F}}_k}{{\bf{v}}_r}} \right|}^2} + \delta _k^2}}}.
\end{equation}\label{eq14}
Similarly, $R_k^{\left( {\rm{p}} \right)} = {\log _2}\left( {1 + {\rm{SINR}}_k^{\left( {\rm{p}} \right)}} \right)$ denotes the achievable rate for the private stream at $k$-th user, ${{\rm{SINR}}_k^{\left( {\rm{c}} \right)}}$ represents the corresponding SINR, i.e.,
\begin{eqnarray}\label{eq15}
{{\rm{SINR}}_k^{\left( {\rm{p}} \right)} = \frac{{{{\left| {{{\bf{F}}_k}{{\bf{v}}_k}} \right|}^2}}}{{\sum\limits_{i \in {\cal K},i \ne k} {{{\left| {{{\bf{F}}_k}{{\bf{v}}_i}} \right|}^2}}  + \chi {{\left| {{{\bf{F}}_k}{{\bf{v}}_r}} \right|}^2} + \delta _k^2}}}.
\end{eqnarray}\label{eq16}
As a key performance indicator, echo SNR is widely used in various radar performance metrics, such as positioning accuracy, detection and false alarm probability. To enhance the received echo signal, a beamformer vector ${{\bf{u}}^H} \in {\mathbb{C}^{1 \times M}}$ is adopted to the radar target, which is denoted as ${{\bf{u}}^H}{{\bf{y}}_r} = {{\bf{u}}^H}{\bf{\hat F}}{{\bf{v}}_r}{s_r} + {{\bf{u}}^H}{{\bf{n}}_r}.$ Subsequently, the echo SNR is given by
\begin{equation}\label{eq17}
{\rm{SN}}{{\rm{R}}_{{\rm{echo}}}}{\rm{ = }}\frac{{{{\bf{u}}^H}{\bf{\hat F}}{{\bf{v}}_r}{\bf{v}}_r^H{{{\bf{\hat F}}}^H}{\bf{u}}}}{{\delta _r^2{{\bf{u}}^H}{\bf{u}}}}.
\end{equation}

\subsection{Problem Formulation}

Under the assumption of perfect channel state information (CSI) at the BS, the goal is to maximize the EE of the proposed system by optimizing the vector of common rate portions ${\bf{c}} = {\left[ {{C_1},\; \cdots ,{C_K}} \right]^{\rm{T}}}$, the transmit beamformer ${{{\bf{v}}}}$, the IRS reflection coefficient matrix ${\bf{\Phi}}$, and the echo receiving beamformer ${\bf{u}}$. This optimization problem can thus be formulated as
\begin{align}
&\mathop {{\rm{\quad\quad\quad\quad\quad\quad}}\max {\rm{ \quad }}\quad \quad\quad \quad \eta }\limits_{{{{\bf{c}}}},{{\bf{v_{\rm{c}}}}},{\left\{ {{{\bf{v}}_k}} \right\}_{k \in \mathcal{K}}},{{\bf{v_{\rm{r}}}}},{\bf{\Phi }},{\bf{u}}}\label{eq18}\\
&{\rm{s}}{\rm{.t}}{\rm{.}}\quad \mathop {\min }\limits_i \left\{ {\left. {R_i^{\left( {\rm{c}} \right)}} \right|i \in {\mathcal{K}}} \right\} \ge \sum\limits_{k \in {\mathcal{K}}} {{C_k}}, \tag{\ref{eq18}{a}} \label{eq18a}\\
&\quad \quad \ C_k \ge  {0},\quad  \forall k \in \mathcal{K},\tag{\ref{eq18}{b}} \label{eq18b}\\
&\quad \quad \ P \le {P_{\max }},\tag{\ref{eq18}{c}} \label{eq18c}\\
&\quad \quad \ {C_k} + {{R_k^{\left( {\rm{p}} \right)}}} \ge R_k^{\left( {{\rm{th}}} \right)},\quad  \forall k \in \mathcal{K},\tag{\ref{eq18}{d}} \label{eq18d}\\
&\quad \quad \ {\rm{SN}}{{\rm{R}}_{{\rm{echo}}}} \ge {\rm{SN}}{{\rm{R}}_{{\rm{th}}}},\tag{\ref{eq18}{e}} \label{eq18e}\\
&\quad \quad \ {\phi _n} \in \left\{ {0,\Delta \phi , \cdots ,\left( {{2^B} - 1} \right)\Delta \phi } \right\},  \forall n \in \mathcal{N},\tag{\ref{eq18}{f}} \label{eq18f}
\end{align}
where (\ref{eq18a}) ensures that the common stream is decoded by all users. (\ref{eq18b}) guarantees that all portions of the common rate remain non-negative. (\ref{eq18c}) enforces a constraint on the total transmit power of the BS. (\ref{eq18d}) defines the communication rate requirement for each user, with $R_k^{\left( {{\rm{th}}} \right)}$ representing the threshold for the total achievable rate of the $k$th user. (\ref{eq18e}) establishes the sensing performance requirement, where ${\rm{SN}}{{\rm{R}}_{{\rm{th}}}}$ refers to the predefined sensing SNR threshold. (\ref{eq18f}) corresponds to the discrete phase shift case, where $B$ represents the number of IRS quantization bits with ${{\Delta \phi  = 2\pi } \mathord{\left/{\vphantom {{\Delta \phi  = 2\pi } {{2^B}}}} \right.\kern-\nulldelimiterspace} {{2^B}}}$.

Since the problem (\ref{eq18}) is formulated and analyzed within a time-varying channel model, this implies that the CSI evolves dynamically over time. The traditional optimization algorithms may struggle to efficiently adjust to such a dynamic environment, leading to severe computational delays and insufficient real-time performance. By continuously learning optimal strategies through interaction with the environment,  the DRL techniques can be able to overcome the shortcomings of traditional methods.

\section{Algorithm Design}

In this section, we first turn (\ref{eq18}) into a Markov decision process (MDP), and then apply PPO algorithm to solve it.

\subsection{MDP}

The action space, the state space and the reward function are described as follows \cite{Ref9}.

\emph{1) \textbf{Action space}}: As described in (\ref{eq18}), the action space comprises five variables, i.e., $\mathcal{A}= \left\{ {{{\bf{v}}_{\rm{c}}},{{\bf{v}}_{\rm{r}}},{\bf{u}},\left\{ {{C_k},{{\bf{v}}_k}} \right\},\left\{ {{\phi _n}} \right\}} \right\}$, where the cardinal number of $\mathcal{A}$ is $\left( {2 \times K + N + 3} \right)$. Specifically, for the $k$-th beamforming vector, it is composed of the power part $\left\| {{{\bf{v}}_k}} \right\|$ and the direction part ${{{\bf{\hat v}}}_k}$, i.e., ${{\bf{v}}_k} = \left\| {{{\bf{v}}_k}} \right\|{{{\bf{\hat v}}}_k}$. Based on the maximum ratio transmission (MRT) and zero-forcing (ZF) schemes, the direction part ${{{\bf{\hat v}}}_k}$ is given by
\begin{equation}\label{eq19}
{{{\bf{\hat v}}}_k} = \left\{ \begin{array}{l}
{{\sum\limits_{i \in {\mathcal{K}}} {{\bf{F}}_i^H} } \mathord{\left/
{\vphantom {{\sum\limits_{i \in {\mathcal{K}}} {{\bf{F}}_i^H} } {\left\| {\sum\limits_{i \in {\mathcal{K}}} {{\bf{F}}_i^H} } \right\|}}} \right.
 \kern-\nulldelimiterspace} {\left\| {\sum\limits_{i \in {\mathcal{K}}} {{\bf{F}}_i^H} } \right\|}},\,{\rm{if \,}}k = 0,\\
{{{{\bf{J}}_k}} \mathord{\left/
 {\vphantom {{{{\bf{J}}_k}} {\left\| {{{\bf{J}}_k}} \right\|}}} \right.
 \kern-\nulldelimiterspace} {\left\| {{{\bf{J}}_k}} \right\|}},\,{\rm{if \,}}k \ne 0,
\end{array} \right.
\end{equation}
where ${{\bf{J}}_k}$ is the $k$-th column of ${\bf{\Xi }} = \left[ {{{\bf{J}}_1}, \cdots ,{{\bf{J}}_K}} \right]$ with ${\bf{\Xi }} = {{\bf{\Gamma }}^H}{\left( {{\bf{\Gamma }}{{\bf{\Gamma }}^H}} \right)^{ - 1}}$ and ${\bf{\Gamma }} = \left[ {{{\bf{F}}_1}, \cdots ,{{\bf{F}}_K}} \right]$. Then, the corresponding power part is given by $\left\| {{{\bf{v}}_k}} \right\| = 0.5\sqrt {{P_{\max }}} \left( {\tanh \left( {\xi _k^{{\rm{POW}}}} \right) + 1} \right)$, where $\tanh \left(  \cdot  \right)$ denotes the hyperbolic tangent function and $\tanh \left( {\xi _k^{{\rm{POW}}}} \right) \in \left[ { - 1,1} \right]$ is employed as an activation function to guarantee the outputs of the deep neural networks (DNNs). Similarly, the achievable rate of the common message $\left\{ {{C_k}} \right\}$ and the phase shift ${\left\{ {{\phi _n}} \right\}}$ are respectively determined via hyperbolic tangent functions, i.e., ${C_k} = 0.5\mathop {\min }\limits_k \left\{ {\left. {R_k^{\left( {\rm{c}} \right)}} \right|k \in {\cal K}} \right\}\cdot\left( {\tanh \left( {\xi _k^{{\rm{COM}}}} \right) + 1} \right)$, ${\phi _n} = 0.5\left( {\tanh \left( {\xi _n^{{\rm{IRS}}}} \right) + 1} \right)\left( {{2^B} - 1} \right)\Delta \phi$, where ${\xi _k^{{\rm{COM}}}}$ and ${\xi _n^{{\rm{IRS}}}}$  are input to the output layer of the neural network. Here, the detailed derivations of ${{\bf{v}}_{\rm{c}}}$, ${{\bf{v}}_{\rm{r}}}$, and ${\bf{u}}$ are omitted due to space limitations.

\textit{2) \textbf{State space}}: To provide the agent with a comprehensive understanding of the environment, the state space includes relevant current channel information, specifically ${{\rm{SINR}}_k^{\left( {\rm{p}} \right)}}$ and ${{\rm{SINR}}_k^{\left( {\rm{c}} \right)}}$. Additionally, the state space comprises the selected action vector $a$ and the instantaneous reward $r$, which intuitively reflects the agent's effectiveness in addressing problem (\ref{eq18}). Therefore, the state space is defined as $\mathcal{S}=\left\{\mathbf{u}, \mathbf{a}, r,\left\{\text{SINR}_{k}^{\text{C}}\right\} ,\left\{ \text{SINR}_{k}^{\text{P}} \right\} \right\}$, with the cardinality of $\mathcal{S}$ given by $\left( {4 \times K + N + 4} \right)$.

\textit{3) \textbf{Reward function}}: Since problem (\ref{eq18}) involves optimization objectives and corresponding constraints, the reward function should incorporate both a reward term and penalty terms for constraint violations. Thus, it includes a reward term as well as penalty terms to address constraint violations, which is given by
\begin{equation}\label{eq20}
r = \eta  \times \left( {{ {\Omega _{{\rm{Com}}}} \times {\Omega _{{\rm{QoS}}}} \times \Omega _{{\rm{Pow}}}} \times {\Omega _{{\rm{echo}}}}} \right),
\end{equation}
where ${{\Omega _{{\rm{Com}}}}}$, ${{\Omega _{{\rm{QoS}}}}}$, ${{\Omega _{{\rm{Pow}}}}}$ and ${{\Omega _{{\rm{echo}}}}}$ respectively denote the penalty coefficients corresponding to the constraints (\ref{eq18a}), (\ref{eq18c}), (\ref{eq18d}) and (\ref{eq18e}), i.e.,
\begin{equation}\label{eq21}
{\Omega _{{\rm{Com}}}} = \left\{ \begin{array}{l}
1,\,\sum\limits_{k \in \mathcal{K}} {{C_k}}  - \mathop {\min }\limits_i \left\{ {\left. {R_i^{\left( {\rm{c}} \right)}} \right|i \in {\mathcal{K}}} \right\} \le 0\\
0,\,\sum\limits_{k \in \mathcal{K}} {{C_k}}  - \mathop {\min }\limits_i \left\{ {\left. {R_i^{\left( {\rm{c}} \right)}} \right|i \in {\mathcal{K}}} \right\}> 0
\end{array} \right.,
\end{equation}
\begin{equation}\label{eq22}
{\Omega _{{\rm{QoS}}}} = \left\{ \begin{array}{l}
1,\,{C_k} + R_k^{\left( {\rm{p}} \right)} - R_k^{\left( {{\rm{th}}} \right)} \ge 0, \quad  \forall k \in \mathcal{K}\\
0,\,{C_k} + R_k^{\left( {\rm{p}} \right)} - R_k^{\left( {{\rm{th}}} \right)} \le 0, \quad  \forall k \in \mathcal{K}
\end{array} \right.,
\end{equation}
\begin{equation}\label{eq23}
{\Omega _{{\rm{Pow}}}} = \left\{ \begin{array}{l}
1,\,P - {P_{\max }} \le 0\\
0,\,P - {P_{\max }} > 0
\end{array} \right.,
\end{equation}
\begin{equation}\label{eq24}
{\Omega _{{\rm{echo}}}} = \left\{ \begin{array}{l}
1,\,{\rm{SN}}{{\rm{R}}_{{\rm{echo}}}} - {\rm{SN}}{{\rm{R}}_{{\rm{th}}}} \ge 0\\
0,\,{\rm{SN}}{{\rm{R}}_{{\rm{echo}}}} - {\rm{SN}}{{\rm{R}}_{{\rm{th}}}} \le 0
\end{array} \right..
\end{equation}

For these constraints, the penalty terms enforce strict adherence to the constraints throughout the optimization process.

\subsection{PPO}

In the PPO framework, a surrogate objective function is given by $J\left( \theta  \right) = {\mathbb{E}_{{\pi _\theta }}}\left[ {{\sigma _t}\left( \theta  \right)A\left( {s,\mathcal{A}} \right)} \right]$, where $\pi_{\theta}$ represents the action-selection policy, parameterized by the DNN with parameters with $\theta$, ${\sigma _t}\left( \theta  \right) = {{{\pi _\theta }\left( {s,{\cal A}} \right)} \mathord{\left/  {\vphantom {{{\pi _\theta }\left( {s,{\cal A}} \right)} {{\pi _{{\theta _{{\rm{old}}}}}}\left( {s,{\cal A}} \right)}}} \right. \kern-\nulldelimiterspace} {{\pi _{{\theta _{{\rm{old}}}}}}\left( {s,{\cal A}} \right)}}$ represents the probability ratio between the current and previous policies, and the advantage function is defined as $A\left( s,\mathcal{A}  \right) =r\left( s_t,\mathcal{A} _t \right) +\mu V_{\pi}\left( s_{t+1} \right) -V_{\pi}\left( s_t \right)$, where $\mu \in \left[ {0,1} \right]$ is the discount factor, ${V_\pi }\left( s \right)$ denotes the state-value function. To conform to the trust region constraint, the proposed approach utilizing PPO achieves the direct maximization of equation (38). Instead, it focuses on optimizing a clipped surrogate objective function, which is formulated as ${J^{{\rm{CLIP}}}}\left( \theta  \right) ={\mathbb{ E}_{{\theta _\pi }}}\left[ {\min \left\{ {{\sigma _t}(\theta )A\left( {s,\mathcal{A}} \right), \omega \left( \theta, s,\mathcal{A}\right)  } \right\}} \right]$, where $\omega \left( \theta, s,\mathcal{A}\right)={\rm{clip}}\left( {{\sigma _t}(\theta ),1 - \epsilon ,1 + \epsilon } \right)\cdot A\left( {s,\mathcal{A}} \right)$, $\epsilon$  is a hyper-parameter that adjusts the clipping fraction of the clipping range. The framework is updated by using the stochastic gradient descent (SGD) method over $\Lambda$ transitions, which is defined as
\begin{equation}\label{eq25}
\theta  = {\theta _{\rm{old}}} - \left( {{1 \mathord{\left/ {\vphantom {1 \Lambda }} \right. \kern-\nulldelimiterspace} \Lambda }} \right) \times \sum\limits_{\left( {{s_t},{\mathcal{A}_t},{r_t},{s_{t + 1}}} \right)}^{\Lambda} {{\nabla _\theta }{J^{{{\rm{CLIP}}}}}( \theta  )}.
\end{equation}
For clarity, the proposed PPO-based approach is outlined in detail in Algorithm 1. As noted in \cite{Ref9}, the time complexity of the proposed PPO-based algorithm is largely influenced by the size of the neural network. For Algorithm 1, the time complexity is approximately $\mathcal{O}\left(\sum_{\ell=1}^{L} {\varpi}_{\ell-1} \cdot {\varpi}_{\ell}\right)$, where $L$ denotes the total number of layers, and ${\varpi}_{\ell}$ indicates the number of neurons in the $\ell$-th layer.

\begin{algorithm}[!t]	
	\caption{The Proposed PPO-Based Approach.}
	\label{Algorithm1}
	\SetKwData{Or}{\textbf{or}}
	\DontPrintSemicolon
	\KwIn {Corresponding channels ${\bf{G}}$, ${\bf{h}}_r$, and ${\bf{h}}_k$;}
	\KwOut {$\mathcal{A}= \left\{ {{{\bf{v}}_{\rm{c}}},{{\bf{v}}_{\rm{r}}},{\bf{u}},\left\{ {{C_k},{{\bf{v}}_k}} \right\},\left\{ {{\phi _n}} \right\}} \right\}$;}
      initialization: Initial action, parameters of the DNN $\theta$, and experience pool; \\

    \While{for episode = 1 to max episode}{

               Receive initial observation state $s_t$, $t = 0$;

	\For{step t = 1 to max time step}
	{
        	Take action $a_t$ based on the current state $s_t$;\\
            Observe an instant reward $r_t$ according to (\ref{eq20});\\
            Observe the next state $s_{t+1}$;\\
            Store the transition $\left( {{s_t},{a_t},{r_t},{s_{t + 1}}} \right)$ into the experience pool;\\
            Sample $\Lambda$ transitions from the experience pool;\\
            Compute the value of the advantage function $A\left( s,\mathcal{A}\right)$;\\
            Update DNN parameters $\theta$ via SGD with by (\ref{eq25});\\
            Update the new current state ${s_t}={s_{t + 1}}$;
	}	
                                   }
\end{algorithm}

\section{Numerical Results}

This section presents numerical results to illustrate insights from the analyses in previous sections. Unless otherwise specified, the simulation parameters are based on \cite{Ref9} and \cite{Ref10}, i.e., $K = 2$, ${K_{{\rm{IR}}}} = 10$, $B = 2$, ${P_{{\rm{ST}}}}$ = 30 dBm, ${P_{\max }}$ = 20 dBm, $\mu  = 1$, $R_k^{\left( {{\rm{th}}} \right)} = 4$ bit/s/Hz, ${\rm{SN}}{{\rm{R}}_{{\rm{th}}}} = 0$ dB, $\delta _r^2 = \delta _k^2 = - 120$ dBm, $\upsilon_r = 5$ ${{\rm{m}} \mathord{\left/{\vphantom {{\rm{m}} {\rm{s}}}}\right.\kern-\nulldelimiterspace} {\rm{s}}}$, $\upsilon_k = 1$ ${{\rm{m}} \mathord{\left/{\vphantom {{\rm{m}} {\rm{s}}}}\right.\kern-\nulldelimiterspace} {\rm{s}}}$, $\gamma_r = \gamma_k = 0^{\circ}$, ${H_B} = 20$ m, ${H_{\rm{I}}} = 25$ m, ${H_{\rm{R}}} = 25$ m, ${x_{\rm{I}}} = 1$ m, ${x_{\rm{R}}} = 1.5$ m, ${x_{\rm{U}}} = 2$ m, ${y_{\rm{R}}} = 1$ m, ${y_{\rm{I}}} = 2$ m, $d_{\rm{B}} = 0.5 \lambda$, $d_{\rm{I}} = 0.2 \lambda$, and $d_{\rm{U}} = 0.5$ m.

\begin{figure}[!t]
\centering
\includegraphics[scale=0.3]{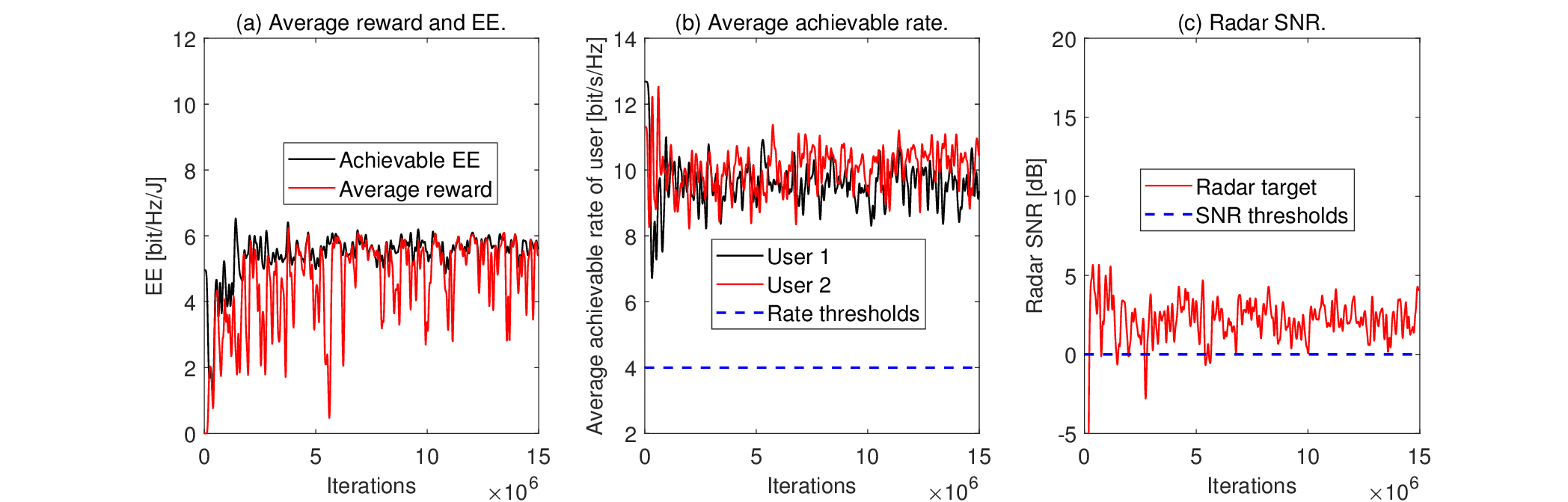}\\
\caption{\label{Fig02} Convergence performance, where ${f_c} = 2.4$ GHz, ${K_{{\rm{BI}}}} = {K_{{\rm{IU}}}}= 10$, $\sigma  = 20$ ${{\rm{m}}^{\rm{2}}}$, $M = 4$, $N = 9$.}
\end{figure}

Fig. \ref{Fig02}(a) shows the convergence of the proposed algorithm, showing that EE and reward stabilize after a finite number of iterations, with the gap between them decreasing over time. This is due to the improved quality of samples in the experience pool as training steps increase. Figs. \ref{Fig02}(b) and \ref{Fig02}(c) display the convergence of achievable rate and radar SNR, respectively, both of which meet the corresponding QoS requirements.

To demonstrate the advantages of the PPO scheme in addressing complex optimization problems, the Random and Greedy approaches are adopted as baseline methods for comparison. As shown in Fig. \ref{Fig04}(a), the proposed PPO scheme achieves the highest performance. This superior performance is attributed to the structured policy updates of the PPO, whereas the Random scheme selects actions uniformly, disregarding the current state and reward structure. In contrast, the Greedy scheme maximizes EE at each step without accounting for long-term outcomes. Fig. \ref{Fig04}(b) shows the EE versus the number of BS antennas for varying the channel conditions and carrier frequency. It is found that the EE declines with increment of the carrier frequency. The reason is that larger carrier frequency results in larger path loss. Specifically, for $M = 8$, the system EE decreases by approximately 67\% as $f_c$ increases from 1.4 GHz to 2.4 GHz. Additionally, double-Rician channels provide higher EE compared to double-Rayleigh channels, suggesting that maintaining a high Rician factor in the fading channel can be beneficial in rich scattering environments, especially when augmented by an IRS.

Fig. \ref{Fig03} demonstrates the EE as a function of the number of IRS elements for varying target RCS. For comparison, we use the SDMA-IRS-assisted ISAC as a benchmark. Results indicate that EE improves with an increasing number of IRS elements in both schemes, underscoring the IRS's contribution to EE enhancement. Furthermore, the proposed algorithm consistently outperforms the benchmark by leveraging SIC to mitigate common stream interference. Notably, the EE increases for both schemes as the target RCS grows, with system EE improving by approximately 50\% as $\sigma$ increases from 10 ${{\rm{m}}^2}$ to 20 ${{\rm{m}}^2}$.

\begin{figure}[!t]
\centering
\includegraphics[scale=0.25]{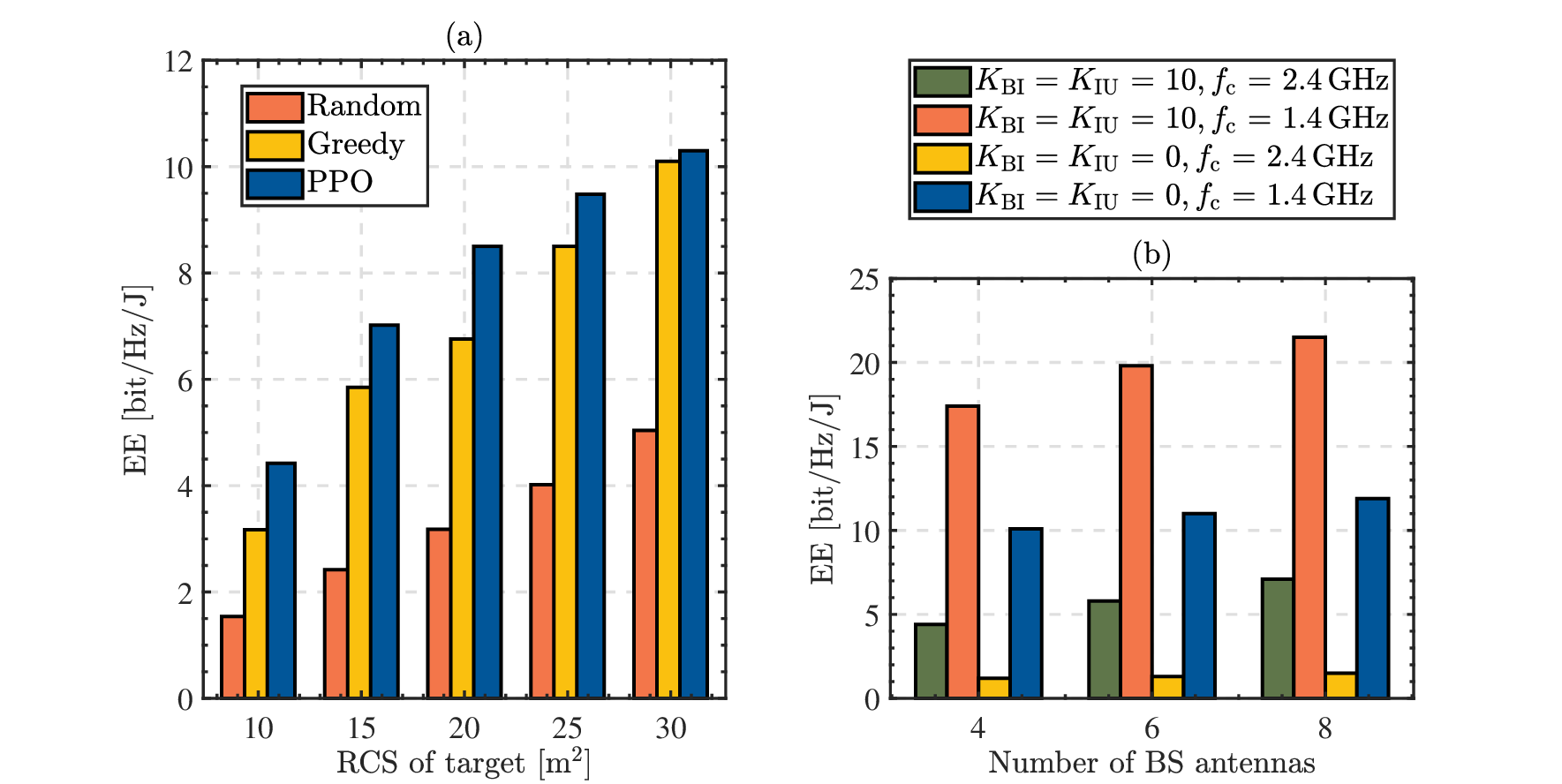}\\
\caption{\label{Fig04} EE comparison of systems with various baseline approaches and channel conditions, where $\sigma  = 10$ ${{\rm{m}}^{\rm{2}}}$, $N = 9$. }
\label{Fig04}
\end{figure}

\begin{figure}[!t]
\centering
\includegraphics[scale=0.3]{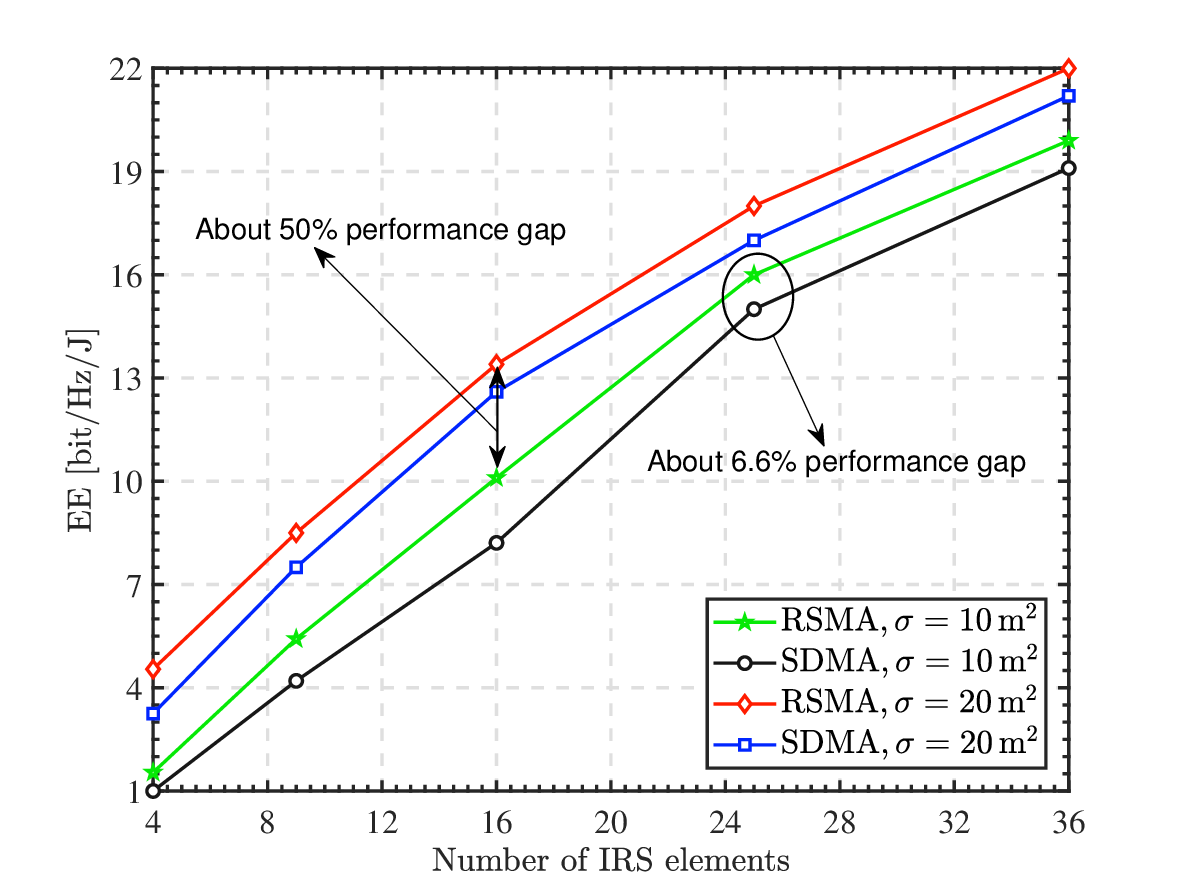}\\
\caption{\label{Fig03} Comparison of EE achieved by RSMA with SDMA, where ${f_c} = 2.4$ GHz, ${K_{{\rm{BI}}}} = {K_{{\rm{IU}}}}= 10$, $M = 4$.}
\label{Fig03}
\end{figure}

\section{Conclusions}

In this paper, we have proposed a 3D geometrical-based channel model, which can be used to accurately characterize the RSMA-IRS-assisted ISAC propagation environments. Then, we have developed the PPO algorithm for maximizing the EE. Finally, the EE was analyzed using the proposed channel model. It has been that (i) fading conditions have strong impact on the EE, (ii) as the number of IRS elements increases, the proposed system achieves higher EE, and (iii) the RSMA-ISAC system outperforms conventional SDMA-ISAC systems.

\end{document}